\documentstyle[prl,aps,multicol,epsf]{revtex}
\begin{document}
\draft
\title{
Magnetization Controlled  Superconductivity\\
in a  Film with Magnetic Dots.}
\author{Igor F.Lyuksyutov$^{a,*}$ and Valery Pokrovsky $^{a,b}$ }
\address{(a) Department of Physics, Texas A\&M University,
College Station, TX 77843-4242 \\
(b) Landau Institute for Theoretical Physics, Moscow, Russia}

\date{\today}
\maketitle
\begin{abstract}
We consider a  superconducting film  with  
Magnetic Dots Array (MDA) placed on it.
Magnetic moments of these dots are supposed to be  
{\it normal} to the film and strong enough
to create vortices in the  superconducting film.
Magnetic interaction between dots is negligible.
Below the superconducting transition 
temperature of the film $T_{2D}$ 
in zero magnetic field the  
MDA with randomly magnetized dots 
produces resistive state of the film.
Paradoxically, in a finite magnetic field
the film with  MDA is superconductive.

\end{abstract}
\begin{multicols}{2}

Recently substantial  progress has been achieved
in the preparation of  magnetic structures connected with  
superconducting films (see  \cite{giv}, \cite{barb}, \cite{shull}). 
In  experiments \cite{barb} a complex of a 
thin ($100 nm$) superconducting film and a {\it single} 
magnetic particle  with
typical dimension  $10-100 nm$ was prepared
by depositing micro-bridge-DC-SQUID onto the particle \cite{barb}.
This type of systems was used 
to study
Macroscopic Quantum Tunnelling (MQT) \cite{barbchu}
of the magnetic moment.
In other type of experiments ( \cite{shull}) 
a superconducting film was placed on the top of a 
periodic array of magnetic dots with typical size 
and distance between dots   $100 nm$.
The magnetoresistance 
of the superconducting film  oscillated
as function of  magnetic flux $\Phi$ through 
the unit cell of MDA \cite{shull}
with the period equal to the magnetic flux quantum $\Phi_0$.
In earlier experiments of this type 
\cite{giv}, \cite{nano}, \cite{ota} the  square lattice  of 
$200 nm$ thick
magnetic particles was mounted upon
$20 nm$ thin superconducting films.
The magnetoresistance and transiton temperature 
of the superconducting film 
displayed periodic variation with
magnetic field with the period 
corresponding to one flux quantum per unit cell.

Magnetic moment of magnetic dots 
can be oriented either parallel or 
normal to the surface of the substrate (superconducting film)
in controllable way \cite{all}, \cite{giv1}. 
In the cited experiments (\cite{giv,barb,shull}) 
magnetic moment of dots
was parallel to the film.  The magnetic flux
from nanomagnets in the direction normal to the 
superconducting film is readily available 
and it can be varied in a broad range 
from a few percent up to many flux quanta 
\cite{barbara}, \cite{giv1} per a dot.  
Depending on the orientation of the dot magnetic
moment and its strength, a single dot can  
create  vortices or pairs of 
vortices, or it can pin the vortices in the 
superconducting film. 
With the help of  nanomagnets one can create 
strong magnetic fields of various
configurations on the surface 
with high precision . 

The vortex-dot pairs 
interact through the magnetic field in vacuum
and in superconducting film. The interaction
between  vortices in the  superconducting film 
grows logarithmically with the distance,
whereas the dipole-dipole
interaction between magnetic moments 
in vacuum decreases with the 
distance. Thus, the magnetic dots bound with 
vortices interact at large distances mainly via vortices
if the distance is still smaller than 
effective screening length $\lambda_{eff}$.
According to Abrikosov 
(\cite{Abr64}, \cite{Abr}),
 $\lambda_{eff} = \lambda^2/d$, where
 $\lambda$ is the London penetration 
depth and $d$ is the film thickness
(see discussion below).

The phase state of the superconducting 
film with the magnetic dot lattice 
depends on the orientation of the  dots magnetic moment
with respect to the film 
and on the magnetic order in the array.
Vortices coupled with magnetic dots create a "frozen" potential 
for individual vortices. 
If the dot magnetic moments are randomly 
oriented, this "frozen" potential fluctuates strongly.
In deep fluctuations  uncoupled
vortices can appear spontaneously.
As a result, the superconducting film may
transit into the resistive state.
However, if magnetic field aligns 
magnetic moments of the dots,
deep random fluctuations of the vortex potential 
are suppressed. 
In a weak enough average magnetic field all
vortices will be pinned by the dots.
Paradoxically, the resistive state exists in zero field,
while non-zero magnetic field restores superconductivity.
Below we consider this phenomenon in more details.

The  amplitude $J$ of the superconducting current 
density per unit area
at distance $r \ll\lambda_{eff}$ from the 
vortex center is given by equation (see e.g. \cite{Abr})
\begin{equation}
 J(\rho ) = { \phi_0  c \over {8 \pi^2 \lambda_{eff} r} }\,\,\, ,
\label{J1} \end{equation}
where $\phi_0$ is the magnetic flux quantum and
$r$ is the distance from the center.
We assume for simplicity that the dot
creates homogeneous magnetic field $H$ 
normal to the surface of superconducting
film and 
localized in a circle of radius $R\ll\lambda_{eff}$. 
The coupling energy $E_d$ between a vortex and the dot is:
\begin{equation}
 E_d = \int^R_0 d r \pi r^2 {1\over c} J(r) H = 
{H \phi_0 R^2  \over {16\pi \lambda_{eff}}} 
= {\epsilon_0 \phi_d \over \phi_0}
\label{Ep} \end{equation}
where $\phi_d =\pi R^2 H$
is the magnetic flux generated by
the magnetic dot and 
$\epsilon_0 = {\phi^2_0 / 16 \pi^2  \lambda_{eff}}$. 
The single vortex energy 
without magnetic field is 
(see e.g. \cite{Abr}):
\begin{equation}
 E_v = 
\epsilon_0 \ln{l \over \xi} 
\label{Ev} \end{equation}
where $l$ is the characteristic cut-off length.
The latter is equal to $\lambda_{eff}$
for a single vortex in the film 
and it is equal to $l =min (a,\lambda_{eff} )$
for a vortex  lattice  with the lattice constant $a$.

It follows from Eqs.(\ref{Ep},\ref{Ev}) that 
the threshold value 
for magnetic flux per magnetic dot necessary to 
create a vortex is equal to 
\begin{equation}
 \phi_{dc} = \phi_0 \ln{l \over \xi} 
\label{phic} \end{equation}
In a typical experimental situation 
the logarithmic factor is not too large.
The magnetic film must be  separated from
the superconducting film by an insulating layer  
(e.g. thin oxide film), 
to avoid the suppression of superconductivity
by the proximity effect. 

Two  vortices with vorticities 
$n_i$ and $n_j$ ($n_i, n_j=\pm 1$) separated by the  distance 
$ r_{ij} \ll \lambda_{eff}$
interact logarithmically
(see e.g. \cite{donhu})
\begin{equation}
 V(r_{ij}) = - {n_i n_j \phi^2_0  \over 
{16 \pi^2 \lambda_{eff}}}\ln{r_{ij} \over \xi} 
\label{Ei} \end{equation}
Eq.(\ref{Ei}) demonstrates explicitly that,
as we discussed earlier, 
the interaction between MD 
with magnetic moments normal to the film via 
attached vortices can be stronger than the 
direct magnetic dipolar interaction 
at large distances. 
Magnetic moment $\cal M$ of the 
magnetic dot with typical size $R$
is of the order of  
${\cal M} \approx H R^3$, where 
$H$ is the magnetic field created by the dot.
The dipolar magnetic interaction between two dots
at distance $r$ can be estimated as 
$ E_{dip}(r) ={ \phi^2_0  R^2 / r^3} $,
where we have used $H R^2 \approx \phi_0$.
Comparing this energy with Eq.(\ref{Ei}) we find 
that the dipolar interaction between MD 
is negligible when  $a\gg R{(\lambda_{eff}/R)}^{1/3}$.
The interaction between dots mediated 
by vortices changes substantially the
magnetization curve when the film is cooled 
below the superconducting transition temperature.
This statement
can be verified experimentally. 

Due to large potential barriers for spin reversal
the magnetic moments in the 
MDA can be  oriented randomly.
Let thethe moments  be large enough 
to create vortices at $T=0$. The vortices
coupled to randomly oriented magnetic moments 
serve as  sources of the  frozen random potential 
$V({\bf r})$ for a single  vortex:
\begin{equation}
 V({\bf r}) = - \sum_{{\bf r}_j} n_j\epsilon_0
K_0({\vert {\bf r}-{\bf r}_{j}\vert \over\lambda_{eff}})
\label{V} \end{equation}
where $n_j$ is the vorticity, i.e. the random variable 
taking values $\pm 1$, and 
$K_0(x)$ is the zero-order MacDonald function
(see e.g. \cite{ryz}). 
For the square MD lattice with the lattice constant $a$
the mean square fluctuation  
of this potential reads:
\begin{eqnarray}
\langle V^2( {\bf r})\rangle 
& = & \epsilon^2_0\sum_{{\bf r}_j,{\bf r}_i} 
\langle n_i n_j\rangle 
K_0({\vert {\bf r}-{\bf r}_{i}\vert  \over \lambda_{eff}})
K_0({\vert {\bf r}-{\bf r}_{i}\vert  \over \lambda_{eff}})\nonumber\\
&=&\epsilon^2_0 \sum_{{\bf r}_j}
K^2_0({\vert{\bf r}_{j}\vert\over \lambda_{eff}})
=\epsilon^2_0 {\lambda^2_{eff}\over a^2}\int_{{\bf r}}K^2_0(r)\nonumber\\
&+&\epsilon^2_0 {\lambda^2_{eff}\over a^2}\sum_{{\bf m}\neq {\bf 0}}
\int_{{\bf r}}K^2_0(r)
\exp(2\pi i {\lambda_{eff}\over a}{\bf r}\cdot{\bf m})
\label{V2} \end{eqnarray}
We assumed uncorrelated vorticities $n_i$, i.e. 
$\langle n_i n_j\rangle = 
\delta_{i,j}$ and 
introduced the vector $\bf m$ with integer 
coordinates  $m_x, m_y$.
Eqn.\ref{V2} allows to find a typical variation 
(depth of potential valleys) in the frozen potential:
\begin{equation}
\Delta V(\lambda_{eff},a)\equiv \sqrt{\langle V^2( {\bf r})\rangle} 
 = \sqrt{\pi/2} \epsilon_0 \lambda_{eff}/a 
\label{V3}
\end{equation}
Fluctuations  with  such energy have maximal allowed 
linear size $r = \lambda_{eff}$. The magnitude 
$\Delta V(\lambda_{eff}, a)$ of a typical potential
fluctuation is increased linearly with $\lambda_{eff}$.
The energy of a single vortex 
grows with  $\lambda_{eff}$ only logarithmically
\cite{nel,donhu,halp,min}:
\begin{equation} 
U_s(\lambda_{eff}) =  \epsilon_0 
\ln(\lambda_{eff}/\xi) +  \epsilon_{core}
\label{U}
\end{equation}
where $\epsilon_{core} \sim \epsilon_0$ 
is the energy of the  vortex core.
Comparing the vortex energy Eq. \ref{U} with
the typical fluctuation Eq. \ref{V3} of the random 
potential, we see that the creation of 
vortex is favorable at $a < \lambda_{eff}$.
It means that, for sufficiently dense MDA or
sufficiently high temperature,
deep valleys in the random 
potential will be filled by unbound vortices.  

When $a >\lambda_{eff}$, fluctuations of 
the random potential can not unbound vortex pairs. 
We discuss now the transition line
$a(\lambda_{eff})$ between the phase with bound 
vortices (superconducting phase) and the vortex plasma (resistive phase).
Let us introduce the MD density  $n_d = a^{-2}$
and its critical value $n^c_d$ at which the transition to
the plasma state occurs at a  given $\lambda_{eff}$.
In the  plasma phase the vortex
density outside of MD is defined by 
the random potential of bound to MD vortices
and by the interaction between free vortices.
In the limit of small vortex density 
$n_v \ll \lambda^{-2}_{eff}$
the average energy gain per vortex is the difference
of the energy of single vortex creation Eqn.\ref{U} 
and the energy gain due to the
random potential Eq. \ref{V3}.
The repulsion between vortices of the same sign
filling a typical random potential valley
increases the energy per vortex by the value
proportional to the  $\epsilon_0\lambda^2_{eff} n_v $ 
(neglecting the logarithmic factor).
Thus, the mean field equation for  energy density has the form: 

\begin{equation}
E_v  =
n_v(\epsilon_0 \ln({\lambda_{eff}\over\xi})  
+ \epsilon_{core}
-\sqrt{\pi\over 2}\epsilon_0{\lambda_{eff} \sqrt{n_d}}) 
+ {B\epsilon_0\over 2}n^2_v \lambda^{2}_{eff}
\label{tr1} 
\end{equation}
where $B$ is a numerical factor of the order of one.
Assuming $ \epsilon_{0}\gg \epsilon_{core}$, 
the equilibrium vortex density following from Eqn.\ref{tr1} is:
\begin{equation}
 n_v\propto {\lambda^{-1}_{eff}}(\sqrt{n_d}-\sqrt{n^{c}_d})
\label{tr2} 
\end{equation}
where $ n^c_d (\lambda_{eff})$ defines the transition line:
\begin{equation}
n_d =  n^c_d \equiv {2\over\pi
{\lambda^{2}_{eff}}}\ln^2 {\lambda_{eff} \over\xi}.
\label{tr3} 
\end{equation}
In the experimental setup the effective penetration 
depth varies with temperature. Thus, the theory predicts
the transition to resistive state in zero field at
some temperature $T_r < T_c$  if the MD density
$n_d$ is smaller than the value 
$\lambda^{-2}_{eff}(0)\ln^2 {\lambda_{eff}(0)\over\xi}$ 
taken at zero temperature.
In a real experiment
the creation of vortex plasma can be very slow at 
$T \ll T_{2D}$ due to potential barriers.
These barriers are of the order
of  $\epsilon_0 $ i.e. of $T_{2D}$. 
We expect that in zero external magnetic field,
far away from the transition line 
Eqn.\ref{tr3}, the densities of vortices 
of both signs in the vortex plasma are  
the same $n_v \approx a^{-1}\lambda^{-1}_{eff}$. 

Qualitatively 
one can realize the influence of random
distribution of vorticities as an 
effective random field of the 
strength ${\cal F} = \sqrt{\pi/2}\epsilon_0 /a $
which acts on vortices. 
The problem of diffusion of a classical particle 
(vortex in our case) in a random field type potential
have been studied in \cite{rf1,rf2}. Dimensionality $D=2$
is marginal for many types of random field 
type disorder. In our case the screening  effectively
stops the growth of the random potential at distances of the order
of $min(a,\lambda_{eff})$. We do not expect 
any  measurable effects in the 
transport properties due to low dimensionality.

Resistivity in the vortex plasma state is defined by equation 
(see e.g. \cite{min,tink}):
\begin{equation}
\rho = 2 \rho_n n_v \xi^{2} \approx   \rho_n {\xi^2\over a\lambda_{eff}}
\label{res} 
\end{equation}
where $\rho_n$ is the resistivity in the normal state.

In the absence of MDA in zero field the superconducting
film undergoes the Berezinskii-Kosterlitz-Thouless 
transition \cite{berez,KT} 
with  the transition temperature $T_{2D} = \epsilon_0 /2$.  
$T_{2D}$ defines characteristic temperature range in
this problem. The Curie temperature $T_{C}$ for magnetic dots 
can be easily choosen in such a way that 
$T_{2D} \ll T_{C}$.
When MDA system is cooled below the Curie point in zero field,
the magnetic moments of the dots are oriented randomly.
In turn, the vortices in a superconducting 
film are also  oriented randomly.
We have shown that this random field can  
fluctuate in space so strongly
that the process of spontaneous vortex-antivortex pair
formation is allowed, leading to  resistive state
of the superconducting film in zero field
at zero temperature. 
Applying an external magnetic field, one can order 
the magnetic moments of the dots. 
In the magnetized
phase of the MDA all vortices bound with dots 
are of the same sign. The periodic array of vortices
which are bound by MD or  created by external magnetic field
produce  strong pinning of the  vortex lattice
and, paradoxically, induces superconducting state with
enhanced critical current. Such magnetized
state of the dots (which is metastable in zero field) 
can exist in zero magnetic field due to large barriers for the 
dot magnetization reversal. 
It vanishes only after annealing.

In conclusion, we have shown that the interaction 
between magnetic dots and superconducting vortices
leads to new types of states and transitions 
between them. We believe that this new kind of interplay
between superconductivity and magnetism opens
broad prospects for fundamental study and applications.

This work was partly supported by the grants
NSF DMR-97-05182 and THECB ARP 010366-003 and
by Minist\`ere de l'Enseignement Sup\'erieur et de la Recherche. 
It is a pleasure to acknowledge discussions with 
B.Barbara, D.Givord, I.Schuller,  D. Naugle and  P.Molho.
I.L.  is  thankful to P.Molho for kind hospitality  
extended to him during his stay
at Laboratoire de Magnetisme Louis Neel in Grenoble where 
part of this work has been done.

\end{multicols}
\end{document}